\documentclass[preprint,superscriptaddress, prl]{revtex4}%
\usepackage{graphicx}%
\usepackage{amsmath}%
\usepackage{amsfonts}%
\usepackage{amssymb}
\providecommand{\U}[1]{\protect\rule{.1in}{.1in}}
\begin{document}
\title{Cyclic Cosmology, Conformal Symmetry and \\the Metastability of the Higgs}
\author{Itzhak Bars}
\affiliation{Department of Physics and Astronomy, University of Southern California, Los
Angeles, CA 90089-0484 USA}
\author{Paul J. Steinhardt}
\affiliation{California Institute of Technology, Pasadena, CA 91125}
\affiliation{Department of Physics and Princeton Center for Theoretical Physics, Princeton
University, Princeton, NJ 08544, USA}
\author{Neil Turok}
\affiliation{Perimeter Institute for Theoretical Physics, Waterloo, ON N2L 2Y5, Canada}

\begin{abstract}
Recent measurements at the LHC suggest that the current Higgs vacuum could be
metastable with a modest barrier (height $(10^{10- 12}~\mathrm{GeV})^{4}$)
separating it from a ground state with negative vacuum density of order the
Planck scale. We note that metastability is problematic for big bang to end
one cycle, bounce, and begin the next. In this paper, motivated by the
approximate scaling symmetry of the standard model of particle physics and the
primordial large-scale structure of the universe, we use our recent
formulation of the Weyl-invariant version of the standard model coupled to
gravity to track the evolution of the Higgs in a regularly bouncing cosmology.
We find a band of solutions in which the Higgs field escapes from the
metastable phase during each big crunch, passes through the bang into an
expanding phase, and returns to the metastable vacuum, cycle after cycle after
cycle. We show that, due to the effect of the Higgs, the infinitely cycling
universe is geodesically complete, in contrast to inflation.

\end{abstract}

\maketitle

The recent discovery at the Large Hadron Collider of a Higgs-like particle
with mass 125-126 GeV \cite{CMS,ATLAS}, combined with measurements of the top
quark mass \cite{CMS:2012fya}, implies that the electroweak Higgs vacuum may
be metastable and only maintained by a modest energy barrier of height
$(10^{10-12}\,GeV)^{4}$ that is well below the Planck density. The conclusion
\cite{Degrassi} is based on computing the running of the standard model Higgs
quartic coupling $\lambda$ and finding that it switches from positive to
negative when the expectation value of the Higgs field $h$ exceeds $10^{10-
12}\,GeV$, under the strong assumption that there is no new physics at
energies less than the Planck scale that significantly alters the predictions
of the standard model. Since there is no evidence at present for disruptive
new physics, we wish to take the metastability seriously and consider its
cosmological implications. The Higgs effective potential is shown in Fig.~1 in
a series of insets that show the potential on progressively smaller scales
(energy density and field are expressed in Planck units).

%%%%%%%%%%%%%
One consequence is that the current phase of the universe, dominated by dark
energy and characterized by accelerated expansion, has a finite lifetime
before decaying to a contracting phase with large negative potential energy
density. There are even more significant consequences for the early universe.
The metastability of the Higgs causes serious problems for big bang
inflationary cosmology~\cite{Espinosa}. If the Higgs is not trapped within the
barrier when the universe first emerges from the the big bang, it will rapidly
evolve to a state with very negative potential energy density, causing the
universe to contract. However, as is apparent from Fig.~1, the barrier height
is so low, the true vacuum so negative (of order the Planck density), and the
metastable field range so narrow compared to the Planck scale that the
likelihood of being trapped is tiny. As for inflation \cite{Guth}, matters are
worse. A generic problem for inflation is that it requires unlikely initial
conditions in order to take hold: namely, the inflaton field must be smooth
and dominant over more than a Hubble volume \cite{Inflahub}. Now, in addition
to the inflaton field being smooth, the Higgs must be trapped over that same
volume or else its negative potential energy density will overwhelm the
inflationary potential energy density, preventing inflation from occurring. A
metastable Higgs thereby makes inflation more improbable. Even if inflation
does begin, de Sitter fluctuations will tend to kick the Higgs field over the
barrier if inflation begins at sufficiently high energies~\cite{Espinosa}.
This effect can terminate inflation at any time, well before the last 60
efolds. In sum, the big bang inflationary picture combined with the metastable
Higgs suggests our universe's past is unlikely and its future is precarious.
The observed universe seems even more unlikely if the Higgs vacuum is part of
a complex energy landscape that in some places includes stable Higgs vacua.

%\begin{figure}[tbh]

\begin{center}
\includegraphics[
width=0.75\textwidth ] {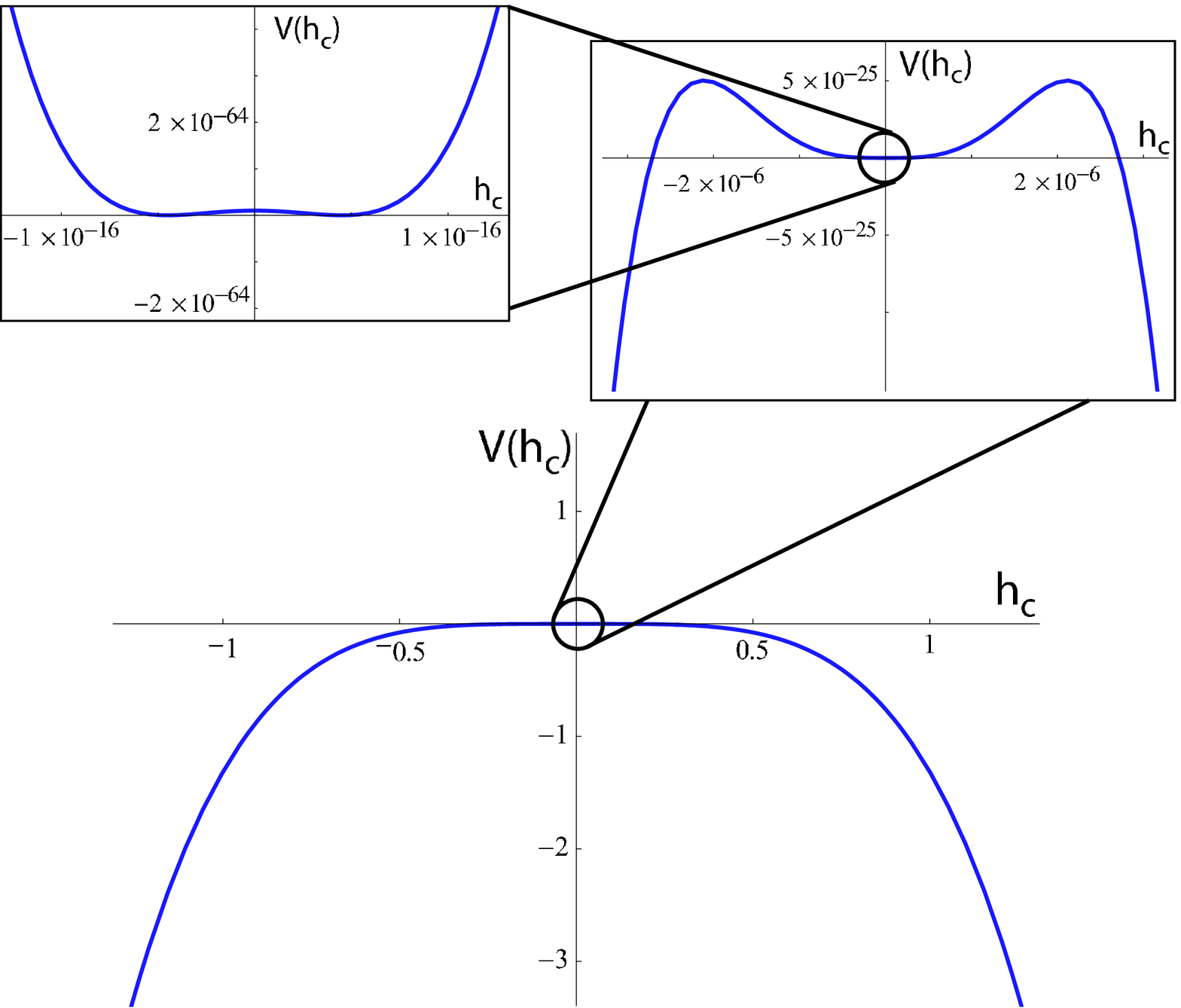}
\end{center}

{\footnotesize Fig.1: A sequence of expanded views of the metastable Higgs
potential for the standard model suggested by recent LHC data. The Higgs field
$|h_{c}|$ and its potential energy density $V(h_{c})$ values are expressed in
Planck units. The bottom figure shows that most of the Higgs field range
corresponds to large negative energy density. The effective potential for
$|h_{c}|>1$ is subject to quantum gravity corrections, so the shape beyond
$|h_{c}|=1$ is unknown. The middle inset (upper right) shows the energy
barrier whose height is nearly 25 orders of magnitude below the Planck
density. The final inset (upper left) shows the current (spontaneous symmetry
breaking) vacuum.}
%\end{figure}

By contrast, a metastable Higgs fits cyclic cosmology~\cite{cyclic1,cyclic2}
to a tee. The current vacuum is \textit{required} to be metastable (or
long-lived unstable), according to the cyclic picture, in order for the
current phase of accelerated expansion to end and for a big crunch/big bang
transition to occur that enables a new cycle to begin. So, it is essential
that there exist scalar fields that can tunnel (or slowly roll) from the
current vacuum with positive potential density to a phase where the potential
energy density is negative and steeply decreasing as the magnitude of the
field grows. The negative potential energy density triggers a reversal from
expansion to contraction that continues as the field rolls downhill. For the
cyclic model, this behavior would not only have to be part of our future, but
also part of our past, describing the period leading up to the most recent
bounce, a.k.a. the big bang.

Hence, a metastable Higgs could play an all-important role in cosmology that
was not anticipated previously. To develop this idea, we constructed a
theoretical formulation \cite{NewestBST} that can incorporate all known
physics and track the evolution of the Higgs through the big bounce. In
constructing this formulation, we were guided by a basic principle that
appears to pervade physics on the very smallest and very largest scales:
scaling symmetry. On the micro-scale, the standard model of particle physics
has a striking scaling symmetry if the small Higgs mass term ($10^{-17}$ in
Planck units) is omitted. This suggests that fundamental physics is
conformally invariant and the desired mass term may emerge from the
expectation value of another scalar field. On the cosmic scale, the Planck
satellite~\cite{Ade:2013rta} has shown the universe to be remarkably uniform
and simple with nearly scale-invariant fluctuations on the largest observable
scales. Together, these observations motivate us to consider Higgs models that
incorporate scale symmetry from the start, including gravity: that is,
Weyl-invariant actions that match phenomenology at the low energies probed by
accelerators \cite{NewestBST}.

A key advantage of these theories for cosmology, as discussed in
Refs.~\cite{NewestBST, cyclicBCT,Bars:2011aa,Bars:2012mt}, is that they have
classical solutions that make it possible to trace their complete evolution
through big crunch/big bang transitions. The completion introduces a period
between big crunch and big bang during which, in the classical, low-energy
description, the coefficient of the Ricci scalar in the gravitational action
changes sign. This brief, intermediate `antigravity' phase is somewhat
analogous to the propagation of a virtual particle within a scattering
amplitude describing incoming and outgoing on-shell particles. In our case,
the incoming collapsing phase and the outgoing expanding phase both involve
`normal' Einstein gravity. We have shown that, in appropriate conformal
gauges, the classical evolution across such a bounce is well-defined and
essentially unique.

In this paper, we explore the question of whether there exist cyclic solutions
that will return the Higgs to its metastable vacuum after each big crunch/big
bang transition. This is not obvious if the Higgs field in the current phase
lies in a shallow potential well, separated by a small barrier (as compared to
the Planck scale) from a very deep negative minimum of Planck scale depth
\cite{Degrassi}. One can imagine that the Higgs field would pop out of the
metastable vacuum during the crunch and never find its way back again in the
next cycle, in which case a metastable Higgs would be incompatible with a
cyclic universe.

For our analysis, we use a Weyl-invariant action $S=\int d^{4}x\mathcal{L}(x)
$ that describes gravity and the standard model
\begin{equation}
\mathcal{L}\left(  x\right)  =\sqrt{-g}\left[
\begin{array}
[c]{c}%
\frac{1}{12}\left(  \phi^{2}-2H^{\dagger}H\right)  R\left(  g\right) \\
+g^{\mu\nu}\left(  \frac{1}{2}\partial_{\mu}\phi\partial_{\nu}\phi-D_{\mu
}H^{\dagger}D_{\nu}H\right) \\
-\left(  \frac{\lambda}{4}\left(  H^{\dagger}H-\omega^{2}\phi^{2}\right)
^{2}+\frac{\lambda^{\prime}}{4}\phi^{4}\right) \\
+L_{\text{SM}}\left(
\begin{array}
[c]{c}%
\text{{\small quarks, leptons , gauge bosons,}}\\
\text{{\small Yukawa~couplings to~}}{\small H,~}\text{{\small dark~matter.}}%
\end{array}
\right)
\end{array}
\right]  \label{action1}%
\end{equation}
Here $L_{SM}$ represents the standard model Lagrangian except for the kinetic
and self interaction terms of the Higgs doublet $H$, which are explicitly
written in the first three lines of Eq.(\ref{action1}). The additional scalar
field $\phi$ is a singlet under $SU(2)\times U(1)$, and, therefore, it cannot
couple to the standard model fields, except for the Higgs, as indicated on the
third line, where $\omega$ is a small parameter ($10^{-17}$ in Planck units)
that determines the Higgs vacuum expectation value and the Higgs mass.
Neutrino masses and simple models of the dark matter may be included through
rather modest extensions involving gauge singlet fields. Both $\phi$ and $H$
are conformally coupled scalars, with the special coefficient $1/12$ required
by the local Weyl symmetry. The action is invariant under Weyl rescaling with
an arbitrary local function $\Omega(x)$ as follows:
\begin{equation}%
\begin{array}
[c]{c}%
g_{\mu\nu}\rightarrow\Omega^{-2}g_{\mu\nu},~\phi\rightarrow\Omega
\phi,\;H\rightarrow\Omega H,\;\\
\psi_{q,l}\rightarrow\Omega^{3/2}\psi_{q,l},\;A_{\mu}^{\gamma,W,Z,g}%
\rightarrow\Omega^{0}A_{\mu}^{\gamma,W,Z,g},
\end{array}
\label{WeylRescaling}%
\end{equation}
Any function that depends only on the ratio $H/\phi$ or $\left(  \det\left(
-g\right)  \right)  ^{1/8}H,$ or $\left(  \det\left(  -g\right)  \right)
^{1/8}\phi,$ is Weyl-invariant.

Note the relative minus sign between $\phi$ and $H$ kinetic energy terms and
couplings to the Ricci scalar $R$. $H$ is the physical scalar field
corresponding to the Higgs, so in the low energy theory there is no choice
about its canonically normalized kinetic energy term, and, then, conformal
symmetry fixes its coupling to $R$. Then, the coupling to the Ricci scalar
must be opposite for $\phi$ in order to obtain the proper positive overall
coefficient of the Ricci scalar in Eq.~(\ref{action1}); with this choice, its
kinetic energy must be opposite as well to maintain conformal invariance. At
first sight, $\phi$ appears to be a ghost. However, this is an illusion, as
can be demonstrated by choosing a Weyl gauge $\Omega(x)$ where $\phi$ is
constant throughout spacetime so that it is eliminated as a physical degree of
freedom. In this gauge (referred to as $c$-gauge in \cite{Bars:2011aa})
$\phi\left(  x\right)  \rightarrow\phi_{0}$ we can express the physically
important dimensionful parameters (the Newton constant $G$, the cosmological
constant $\Lambda$, and the electro-weak scale $v$) as:
\begin{equation}
\frac{1}{16\pi G}=\frac{\phi_{0}^{2}}{12},\;\frac{\Lambda}{16\pi G}=\frac
{1}{4}\lambda^{\prime}\phi_{0}^{4},\;H_{0}^{\dagger}H_{0}=\omega^{2}\phi
_{0}^{2}\equiv\frac{v^{2}}{2}. \label{dimensionfulConstants}%
\end{equation}

The original action, Eq.~(\ref{action1}), determines the conformally-invariant
effective action for the relevant cosmological degrees of freedom for a
homogeneous and isotropic Friedmann-Robertson-Walker (FRW)
universe\cite{Bars:2012mt}:
\begin{equation}
\int d\tau\left(
\begin{array}
[c]{c}%
-\frac{1}{2e}[\left(  \partial_{\tau}(a\phi)\right)  ^{2}+\left(
\partial_{\tau}(ah)\right)  ^{2}]\\
-e\left[  a^{4}V(\phi,h)+\rho_{r}+C\sqrt{\rho_{r}}a^{2}h^{2}+\mathcal{K}%
(\phi^{2}-h^{2})a^{2}\right]
\end{array}
\right)  \label{cosmoAction}%
\end{equation}
where $\tau$ is conformal time, $e$ is the lapse function, $C$ is a
dimensionless constant, $\mathcal{K}$ is the spatial curvature, and the
homogeneous function of degree four, $V(t \phi, \, th) = t^{4} V(\phi,\, h)$
describes the Higgs potential. Here we treat the gauge bosons and fermions as
a radiation fluid at temperature $T$, inducing a term of the form
$T^{2}H^{\dagger}H \sim\sqrt{\rho_{r}} a^{2} h^{2}$ in the effective potential
for the Higgs field, where $\rho_{r}/a_{E}^{4} \propto T^{4}$ is the radiation
density in Einstein frame and $\rho_{r}$ is a constant.

The classical equations following from Eq.~(\ref{cosmoAction}) can be analyzed
in various conformal gauges (c-gauge, E-gauge, $\gamma$-gauge) as described in
Ref.~\cite{Bars:2011aa}. In each gauge we label the fields with a
corresponding subscript $\left(  a_{c},\phi_{c},h_{c}\right)  $ or $\left(
a_{E},\phi_{E},h_{E}\right)  $ or $\left(  a_{\gamma},\phi_{\gamma},h_{\gamma
}\right)  $. In the c-gauge already described, the conformal gauge freedom in
(\ref{WeylRescaling}) is used to set $\phi_{c}=\phi_{0}=1$ in Planck units,
eliminating the $\phi$ degree of freedom. In the Einstein gauge, the
coefficient of the Ricci scalar in (\ref{action1}) is set to a constant
$\frac{1}{12}(\phi_{E}^{2}-h_{E}^{2})=\frac{1}{2}$, reducing $\left(
\phi,h\right)  $ to a single scalar degree of freedom. Finally, in the
unimodular or $\gamma$-gauge the determinant of the metric is set equal to
minus one, or $a_{\gamma}=1$. In this gauge there clearly is no cosmological
singularity, while all the dynamics including the expansion of the universe is
represented smoothly by the fields $\phi_{\gamma}$ and $h_{\gamma}$. The
cosmological evolution may be studied in any gauge, but for the purposes of
analyzing and interpreting the solutions it is often useful to translate the
results into gauge-invariant quantities whose physical meaning is clear in
some particular gauge. One such quantity is $h/\phi=h_{c}/1=h_{E}/\phi
_{E}=h_{\gamma}/\phi_{\gamma}$, which represents the magnitude of the Higgs
field in Planck units in $c$-gauge ($h_{c}$). Another is \cite{Bars:2011aa}
$\chi=\frac{1}{6}(-g)^{\frac{1}{4}}(\phi^{2}-h^{2})= \frac{1}{6}(1-h_{c}^{2})=
\frac{1}{6}(\phi_{\gamma}^{2}-h_{\gamma}^{2})=a_{E}^{2}$sign$\left(
\chi\right)  $, which represents the square of the scale factor in $E$-gauge,
$a_{E}^{2}=|\chi|=\frac{1}{6}\left\vert \phi_{\gamma}^{2}-h_{\gamma}%
^{2}\right\vert ;$ note that sign$\left(  \chi\right)  $ is gauge invariant.
Yet a third useful gauge-invariant quantity is $a \phi= a_{c} \cdot1 = 1
\cdot\phi_{\gamma}$.

For finding and exploring bouncing FRW cosmologies, the unimodular $\gamma
$-gauge is most convenient, as discussed in Ref.~\cite{Bars:2011aa}. In the
case that the Higgs potential is of purely quartic form, $V\left(
\phi_{\gamma},h_{\gamma}\right)  =\frac{1}{4}\left(  \lambda h_{\gamma}%
^{4}+\lambda^{\prime}\phi_{\gamma}^{4}\right)  ,$ we have produced complete
analytic solutions \cite{Bars:2012mt} for $\left(  \phi_{\gamma}\left(
\tau\right)  ,h_{\gamma}\left(  \tau\right)  \right)  $ for all values and
signs of $\left(  \lambda,\lambda^{\prime}\right)  ,$ including radiation
$\rho_{r}$, curvature $\mathcal{K}$, and all initial conditions for the fields
$\left(  \phi_{\gamma},h_{\gamma}\right)  $ and their derivatives $\left(
\dot{\phi}_{\gamma},\dot{h}_{\gamma}\right)  .$ (We did not consider the
thermal contribution proportional to $C$ in Eq.~(\ref{cosmoAction}) in
\cite{Bars:2012mt}, but its inclusion is trivial in the same approach.) These
studies yielded all solutions not just some special cases, thus teaching us
how to construct bouncing cosmological spacetimes for all the fields
$\phi,h,g_{\mu\nu}. $
%% This was an invaluable exercise that provided us
%% with new insights
%% and tools for examining
%% cosmological equations.

The realistic Higgs potential analyzed in this paper is a small deformation of
the quartic potential above for which exact analytic solutions were obtained,
so the generic properties of the cosmological solutions are similar. We will
discuss the realistic case below using numerical methods but guided by our
knowledge of the exact solutions so that we know our solutions are generic
rather than based on wishful assumptions about initial conditions.

The evolution of an FRW universe with Higgs field and radiation is represented
in $\gamma$-gauge by the two dynamical quantities $\phi_{\gamma}$ and
$h_{\gamma}$. Using the gauge invariant quantity $|\chi| = a_{E}^{2} =
|\phi_{\gamma}^{2} - h_{\gamma}^{2}|$, one sees that the cosmic singularity in
Einstein frame at $a_{E}=0$ corresponds to crossing the light-cone in the
$\phi_{\gamma}-h_{\gamma}$ plane (see \cite{Bars:2011aa,Bars:2012mt}). In
unimodular $\gamma$-gauge, for which $a_{\gamma}=1$, solutions are smooth
across the light-cone and, hence, can be continued through the big crunch/big
bang transition. Thus, in this context, unimodular gauge and all gauges
smoothly related to it are regarded as good conformal gauges. In contrast, in
$E$-gauge some quantities are singular at $a_{E}=0$. For example, $E$-gauge
assumes that the gauge-invariant quantity $1-h^{2}/\phi^{2}$ is non-negative;
however, our complete set of solutions show that this is not true for generic
initial conditions. Hence, $E$-gauge is a bad gauge choice for studying the
complete evolution of FRW cosmologies.

To study the metastable Higgs, we numerically solve the equations of motion
for the action in Eq.~(\ref{cosmoAction}) using the quantum corrected Higgs
potential:
\begin{equation}
V(\phi,h)\equiv\frac{1}{4} \phi^{4} \left(  \lambda^{\prime}+\lambda(h/\phi)
\left(  \frac{h^{2}}{\phi^{2}}-\omega^{2}\right)  ^{2}\right)  ,
\label{higgspotential}%
\end{equation}
where the factor multiplying $\phi^{4}$ is Weyl-invariant. In the $c$-gauge,
with $\phi_{c}=\phi_{0}=1$ in Planck units, this looks like the familiar Higgs
potential including the quantum corrected running coupling $\lambda\left(
h/\phi_{0}\right)  .$ Then $\omega=\left(  246~\text{GeV}\right)  /\phi_{0}$
gives the Higgs expectation value, and $\frac{1}{4}\lambda^{\prime}\phi
_{0}^{4}$ gives the cosmological constant, both in Planck units in today's
Higgs vacuum.

%\begin{figure}[tbh]

\begin{center}
\includegraphics[
width=0.6\textwidth ] {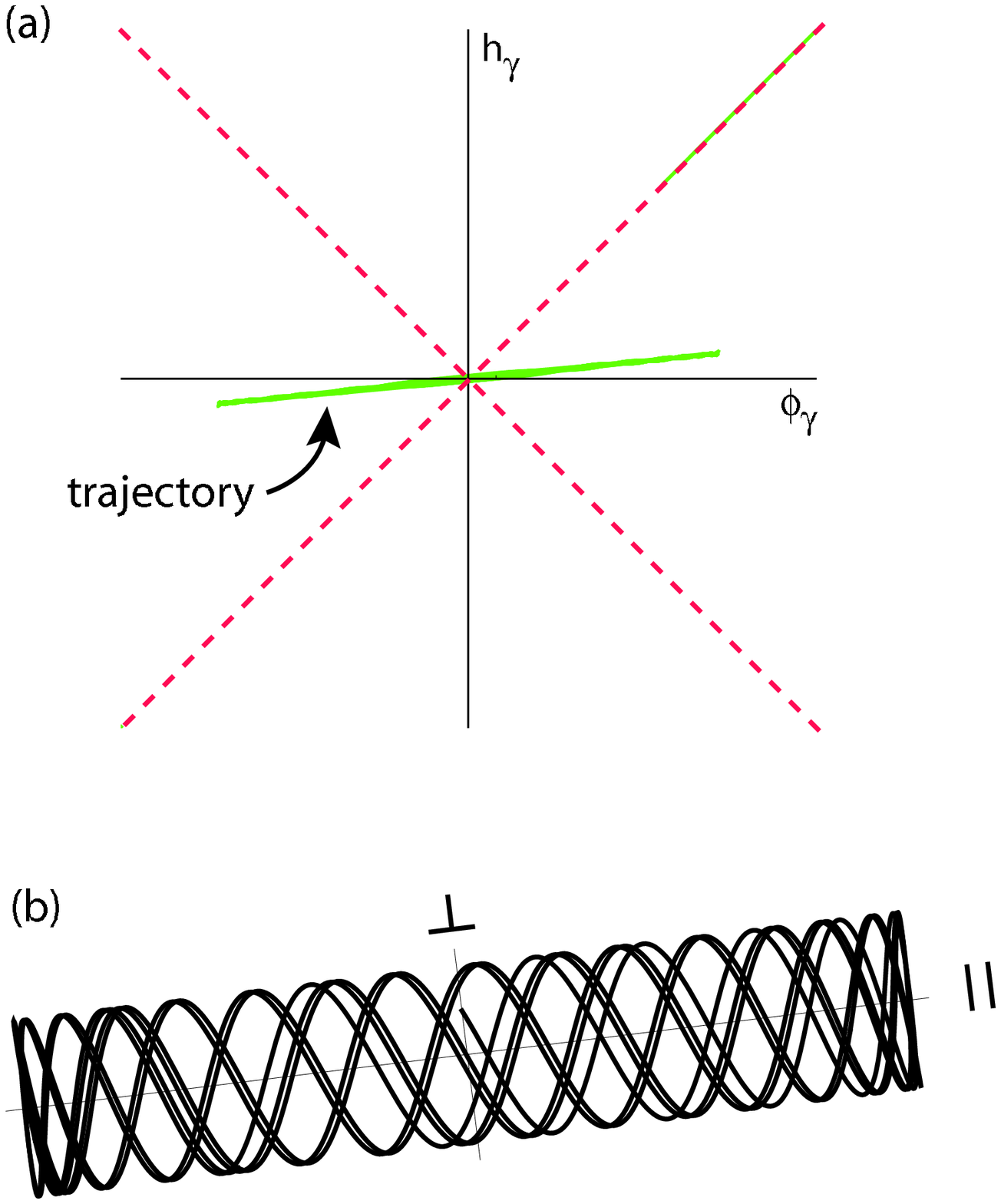}
\end{center}

{\footnotesize Fig.2: Figure (a) shows the back-and-forth trajectory of
$h_{\gamma}$ and $\phi_{\gamma}$ in the $\phi_{\gamma}$-$h_{\gamma}$ plane.
Although the trajectory appears to be a straight slanted line when shown on
this scale, Figure (b) is a blow-up that shows the trajectory to be more
complicated. In this plane, the light-cone like lines correspond to the
singularity in the Einstein frame, $a_{E}^{2}= (1/6)|\phi_{\gamma}^{2} -
h_{\gamma}^{2}|=0$. In the left and right quadrants, the trajectory moves
along an approximately fixed angle corresponding to a fixed Higgs value
(constant $h_{c}=h_{\gamma}/\phi_{\gamma}$), the current vacuum. As the
trajectory moves towards the lightcone, the universe contracts ($a_{E}$
shrinks); crossing the lightcone corresponds to the big crunch/big bang
transition; and moving away from the lightcone corresponds to expansion
(growing $a_{E}$). In Figure (b), the axis perpendicular to the trajectory
($\perp$) has been greatly expanded to show the combination of oscillations
perpendicular and parallel ($\parallel$) to the trajectory, characteristic of
the eternally cycling solution. From this figure it can be seen that cycles
are quasiperiodic: similar but not identical, they explore an invariant torus
in phase space as expected from the KAM theorem.}
%\end{figure}

In order to study cyclic solutions, in this paper we shall artificially take
$\lambda^{\prime}$ to be negative and small compared to all other scales. This
is to mimic an additional effect needed in a cyclic model, where
$\lambda^{\prime}$ would be replaced by a field that rolls or tunnels from
small positive energy density (corresponding to the current dark energy
density) to a negative value to trigger the transformation from expansion to
contraction. This field could even be the Higgs if tunneling is included. That
is, when the Higgs tunnels, it jumps to a state with negative potential energy
density that plays the same role as $\lambda^{\prime}$ in transforming the
universe from expansion to contraction. For the purposes of exhibiting the
stable cyclical behavior of the Higgs, however, the same effect can be
obtained by setting $\lambda^{\prime}<0$, in which case the transformation
occurs when the density of matter and radiation in the current spontaneous
symmetry breaking vacuum falls below $|\lambda^{\prime}|$.

For the running quartic coupling $\lambda(h/\phi)$, we assume the form
computed in Ref.~\cite{Degrassi}. Rather than use the precise result, which
cannot be easily expressed in closed form, we use a simplified expression that
captures the essential features: a metastable, spontaneous symmetry breaking
Higgs vacuum, with a barrier of $\sim$ ($10^{10-12}$~GeV$)^{4}$ separating it
from the true negative energy density vacuum. A simple parameterization that
reproduces the key features in Fig.~1 is:
\begin{equation}
\lambda(h/\phi)=\lambda_{0}\left(  1-\epsilon\,\ln\left(  \frac{h}{\omega\phi
}\right)  ^{2}\right)
\end{equation}
where $\lambda_{0}$ is chosen to fit the observed Higgs mass in today's Higgs
vacuum at $h/\phi=\omega\approx10^{-17}$, and $\epsilon$ is chosen such that
the quartic coupling passes below zero at $h_{c}\approx10^{12}$~GeV. (To avoid
logarithmic singular behavior for our numerical computations, we include small
cutoff parameters inside the log not shown here because the solutions are
insensitive to them.)

Our principal finding is that there exists a continuous band of solutions that
undergo repeated cycles of expansion, contraction, crunch, bang and back to
expansion again in which the Higgs field returns to the metastable Higgs
vacuum during each expansion phase and that these infinitely cycling solutions
are geodesically complete. The band corresponds to solutions whose total Higgs
kinetic plus potential energy density lies in a range that extends from a
little above the barriers in Fig.~1 (second inset) to the local minimum of the
potential corresponding to the current vacuum. As long as the Higgs initial
condition lies in this band after the bang, it returns to the stable band
after each subsequent big bang. It is then trapped within the depression
within the potential barriers (second inset of Fig.~1) and its kinetic energy
red shifts until it settles into a spontaneous symmetry breaking vacuum (third
inset of Fig.~1). Due to the negative cosmological constant ($\lambda^{\prime
}$) the total energy density eventually becomes negative and the evolution
reverses from expansion to contraction. Now the Higgs field kinetic energy
density begins to blue shift until its oscillations grow to the point where
its jumps beyond the barriers and approaches the Planck scale at the big
crunch. After passing through the region with $h_{\gamma}^{2} >\phi_{\gamma
}^{2}$, the process begins again.

The trajectory in the $\phi_{\gamma}$-$h_{\gamma}$ plane is illustrated in
Fig.~2 for the case of no anisotropy. The evolution of $\phi$, $h$ and the
gauge-invariant ratio $h_{c} = h/\phi$ corresponding to the Higgs field value
in $c$-gauge are shown in Fig.~3. The 45 degree lines correspond to $a_{E}^{2}
= |(1/6)(\phi_{\gamma}^{2} - h_{\gamma}^{2})|=0$, a singularity corresponding
to either a big crunch or a big bang. Between the crunch and bang is a brief
intervening period in which $h_{\gamma}^{2} > \phi_{\gamma}^{2}$ and the
coefficient of $R$ in the action (\ref{action1}) changes sign, as discussed in
Ref.~\cite{Bars:2011aa}. Fig.~2 shows that the solution passes without
incident through each crunch/bang transition. As the trajectory passes through
the light-cone-like boundaries in Fig.~2, $|h_{\gamma}/\phi_{\gamma}|$
approaches unity (see the jumps in Fig.~3a), so the Higgs field has popped out
of the metastable vacuum, as anticipated. Then, beginning from the left
quadrant say, the trajectory goes through a period of contraction, passes
through a crunch (the first 45 degree line) and bang (the second 45 degree
line), and enters a period of expansion where it traverses deep into the right
quadrant, corresponding to increasing $a_{E}^{2} \sim\chi\propto\phi_{\gamma
}^{2} - h_{\gamma}^{2}$. During this phase, the Higgs field is observed to
move towards zero and, as the expansion continues, to oscillate and slowly
settle down (due to Hubble red shift) into one of the symmetry breaking vacua,
as discussed above. Due to red shifting, the sum of the (positive) radiation
and Higgs oscillatory energy densities plus the (negative) cosmological
constant term $\lambda^{\prime}\phi_{\gamma}^{4}$ eventually reaches zero. The
Hubble expansion reverses to contraction, the trajectory begins to move
towards the left quadrant, the radiation and Higgs oscillation densities begin
to grow due to blue shifting until the crunch and bounce, and the cycle begins
again. The cycles are not identical, as can be seen from the back and forth
trajectories over several cycles in Fig.~2b and by carefully comparing the
Higgs oscillations from one cycle to the next in Fig.~3a. The classical
equations turn out to obey the assumptions of the Kolmogorov-Arnold-Moser
(KAM) theorem, according to which a weakly nonlinear perturbation of a
classically integrable system, generically deforms but does not remove the
invariant tori in phase space. In our case, the equations are integrable in
the case of no coupling between $h$ and $\phi$, as shown in \cite{Bars:2012mt}%
, and the small coupling between $h$ and $\phi$ is a perturbation. Hence, the
system cycles forever in quasi-periodic fashion and only explores a torus in
phase space that is stable under perturbations, as illustrated in Fig.~2b.

%\begin{figure}[tbh]

\begin{center}
\includegraphics[
width=0.6\textwidth ] {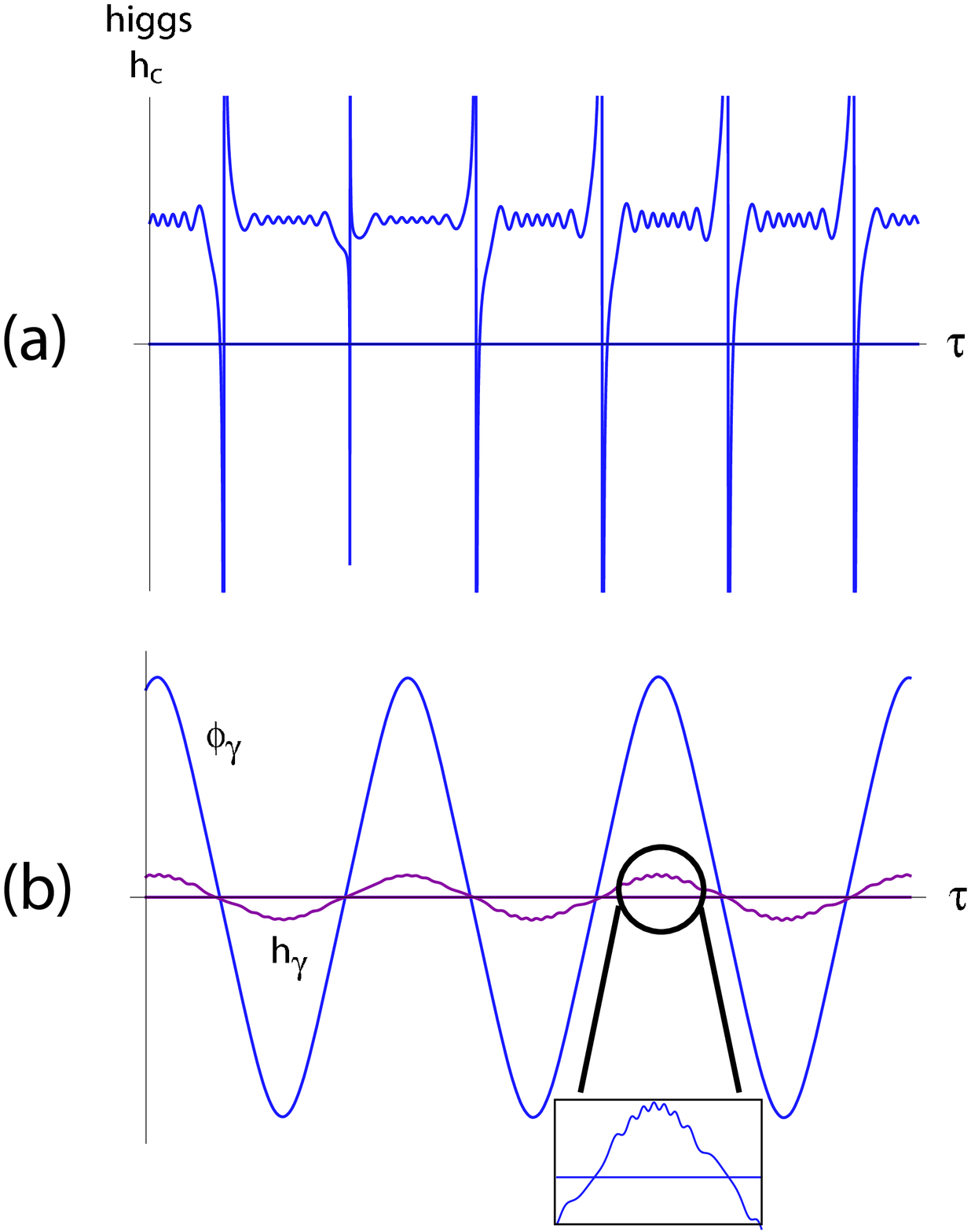}
\end{center}

{\footnotesize Fig.3: Graphs of the evolution of (a) the Higgs field ($h_{c} =
h_{\gamma}/\phi_{\gamma}$), and (b) the scale factor ($a_{\gamma}=
\phi_{\gamma}$) and $h_{\gamma}$, for the trajectory shown in Fig.~2. In (b),
$\phi_{\gamma}$ oscillates smoothly while the evolution of $h_{\gamma}$ has
barely detectable, high frequency oscillations (magnified in the inset)
corresponding to the Higgs field oscillating back and forth in the potential
well of the metastable phase. Note that, in (a), the Higgs oscillates around a
fixed value (the current vacuum) for most of a cycle and then pops to large
values up to the Planck scale during the crunch/bang transition. When the
spikes in (a) cross $\pm1$ (the Planck scale, beyond the range shown here),
the lightcone in Fig.~2a is being crossed, corresponding to a big crunch/big
bang transition where $a_{E}$ vanishes.}
%\end{figure}

Although the figures illustrate the case with no anisotropy, the effects of
anisotropy (combined with radiation) can be easily surmised based on the
results in Ref.~\cite{Bars:2011aa}. Namely, anisotropy only has a significant
role near the big crunch/big bang transition, modifying all passes through the
lightcone shown in Fig.~2a. The paths are distorted so that they travel
precisely through the origin of the $\phi_{\gamma}$- $h_{\gamma}$ plane,
traverse a small loop in the upper or lower quadrants and then exit through
the origin again into the next big bang expansion phase. The size of the loop
depends on the radiation density; higher radiation density results in a
smaller loop \cite{Bars:2011aa}. (To compare cases with and without
anisotropy, see Fig.~1 in \cite{Bars:2011aa}.) This imposes a significant
dynamical behavior on ($h_{\gamma}$,~$\phi_{\gamma}$) where they are both
forced to be zero at the bang, but approach a ratio which is exactly 1,
namely, $h_{c}=h_{\gamma}/\phi_{\gamma}=1$ at the crunch/bang
\cite{Bars:2011aa}. After the bang and far from the singularity, the distorted
path approaches close to the trajectory without anisotropy. So anisotropy only
helps ensure that the Higgs field is trapped in the next phase of expansion.%

%\begin{figure}[tbh]
\begin{center}
\includegraphics[
%natheight=4.181400in, natwidth=5.335900in, height=4.2324in,
%width=5.393in ] {BSTHiggsfig04.eps}
width=0.9\textwidth ] {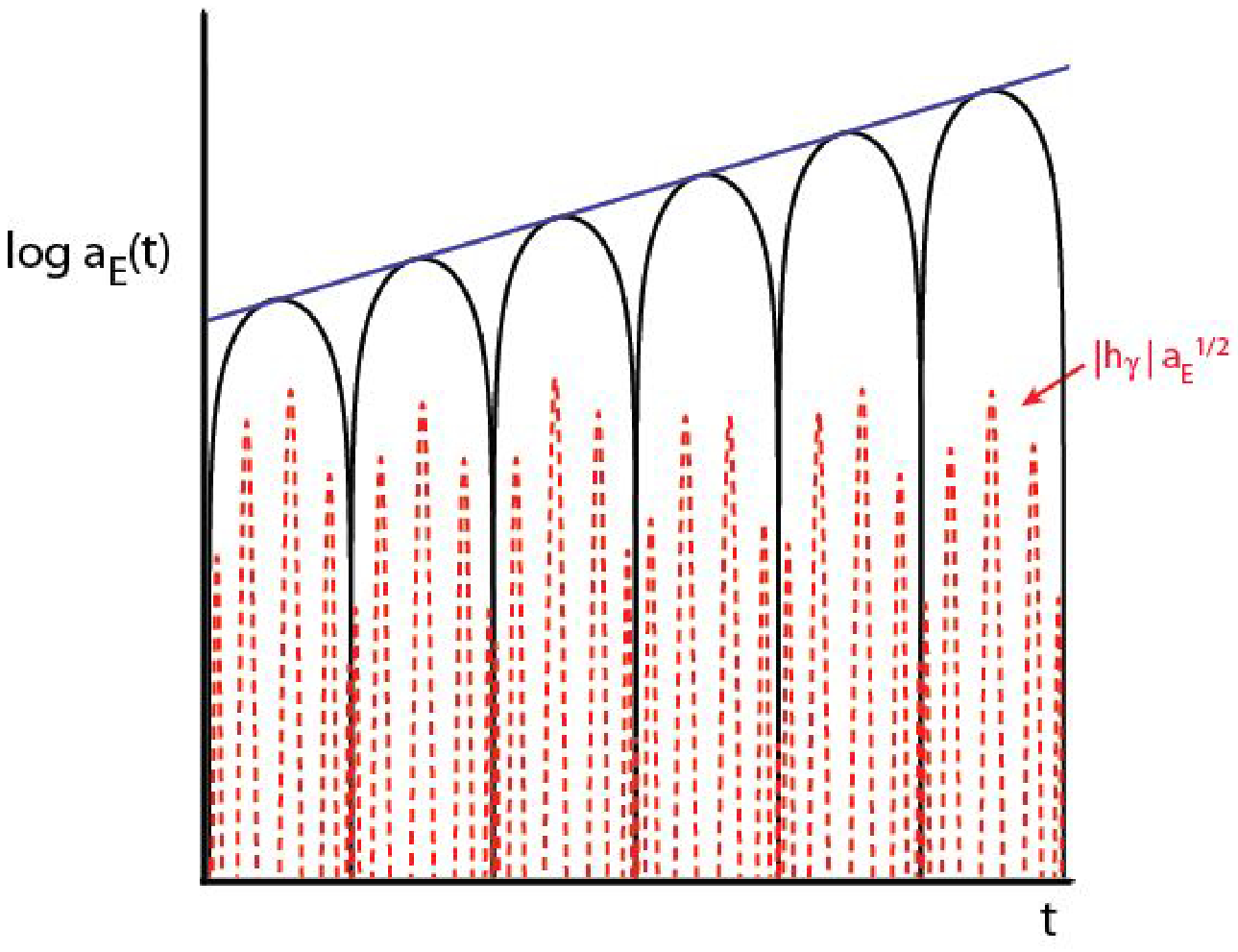}
\end{center}
{\footnotesize Fig.4: Plot of the Einstein frame scale factor $a_{E}$ and
Higgs field value in $\gamma$-gauge $h_{\gamma}$ versus cumulative FRW time,
$t = \int d\tau a_{E}(\tau)$. If the entropy increases by a constant factor
every cycle, then so does the scale factor (solid curve) leading to
exponential growth over many cycles, as indicated by the slanted line. The
Einstein-frame temperature at given cosmic time $t$ is the same as it was a
cycle earlier or will be a cycle later; and the behavior of $|h_{c}|
a_{E}^{1/2}$ (shown in red dashed curve), is the same on average from cycle to
cycle. That is, as the maximum $a_{E}^{2} = \phi_{\gamma}^{2} - h_{\gamma}%
^{2}$ increases from one cycle to the next, the amplitude of the oscillations
in $|h_{c}|= |h_{\gamma}/\phi_{\gamma}|$ decreases. }
%\end{figure}

It is natural to imagine that significant radiation density (parameterized by
$\rho_{r}$) is produced during the bounce that couples to the Higgs. In
Ref.~\cite{Bars:2011aa}, it was shown that radiation produced during the
antigravity phase backreacts by speeding up $\phi_{\gamma}$ and $h_{\gamma}$
such that the universe emerges from the big bang more rapidly than it entered
the big crunch, with the consequence that the scale factor $a_{E}$ grows from
cycle to cycle, as shown in Fig.~4. The duration of each cycle in proper FRW
time is set by the value of the (negative) cosmological constant
$\lambda^{\prime}$, so it does not change. Likewise, the temperature of the
universe in Einstein frame at maximum scale factor does not change from cycle
to cycle. This echoes the scenario envisaged in the cyclic model in
Ref.~\cite{cyclic1}, but, remarkably, here it is accomplished through the
standard model with a metastable Higgs. What does change with each cycle is
the amplitude of the oscillations in the Higgs field around the
symmetry-breaking minimum. As radiation is produced, the parameter $\rho_{r}$
grows, and the Higgs oscillation amplitude decreases.

Finally, we wish to demonstrate the surprising effect the Higgs can have in
enabling cyclic universes to be geodesically complete, in contrast to
inflationary scenarios, including ``eternal inflation." Inflationary scenarios
are well-known to be at most semi-eternal to the future and, hence,
necessarily dependent upon some assumed initial condition \cite{Borde:2001nh}.
For our cyclic solutions, we have already emphasized in \cite{Bars:2011aa} and
here that the solutions may be continued arbitrarily far backward in conformal
time through bounce after bounce. We shall now compare cyclic and inflationary
scenarios using a coordinate-invariant definition of geodesic completeness.

Inflationary scenarios, if they allow for \textit{any} forms of energy other
than inflationary potential energy at early times, generically lead to a
singularity at finite conformal time in the past. In considering the global
structure of inflationary spacetime, it is sometimes considered that the
universe somehow became stuck an arbitrarily long time in the past in a
positive energy false vacuum. In the flat slicing of pure de Sitter spacetime,
the conformal time during inflation then stretches arbitrarily far back into
the past. However, this superficial appearance of completeness is in fact a
coordinate artifact. This may be seen by changing to the closed slicing of de
Sitter spacetime. In this slicing, the universe `bounces' at some finite
conformal time in the past, so that inflation is in fact preceded by a
deflationary phase in which all of inflation's successes during the expanding
de Sitter phase would be precisely undone in a preceding, collapsing de Sitter
phase. So in order to build a successful inflationary scenario, one must
simply ignore the earlier collapsing phase and assume that the universe just
somehow started out in the expanding phase. This is part of the well-known
initial conditions problem of inflation.

Even without changing to the closed slicing, one can identify the problem by
using the following coordinate-invariant definition of geodesic completeness:
that generic timelike geodesics -- the worldlines of massive freely falling
particles -- may be extended arbitrarily far backward in proper time. It is
natural to measure the time along the particle worldline $t_{p}$ in units of
$m^{-1}$ where $m$ is the particle mass, \textit{i.e.}, the magnitude of the
action for the particle:
\begin{equation}
|\mathcal{S}| = \int m dt_{p} = \int d \tau{\frac{m^{2} a_{E}(\tau)^{2}
}{\sqrt{p^{2} +m^{2} a_{E}(\tau)^{2}}}} , \label{propertime}%
\end{equation}
(see, \textit{e.g.}, \cite{cyclicBCT}), where $p$ is the particle's
(conserved) canonical momentum (and $\tau$ is the conformal time, as before).
In the case of an expanding de Sitter epoch stretching all the way back to
zero scale factor, the best possible case for inflation, we have
$a\propto-1/\tau$, where $\tau= -\infty$ corresponds to $a=0$. The integral
converges at the lower limit, meaning that the total proper time experienced
by the particle is finite, so the spacetime is geodesically incomplete. Since
this was the best possible case, it follows that, in the absence of an account
of what preceded inflation, all inflationary scenarios, even `eternal
inflation' scenarios, are geodesically incomplete~\cite{Borde:2001nh}.

In contrast, with the same criterion, all of our Higgs cyclic models are
geodesically complete. All massive particles receive a mass contribution from
the Higgs, and so the quantity $m a_{E}$ in (\ref{propertime}) should be
replaced by $g h_{E} a_{E}$, where $g$ is some coupling constant. Since $ha$
is gauge-invariant, we can replace this $g h_{E} a_{E} =g h_{\gamma}
a_{\gamma} = g h_{\gamma}$, where $h_{\gamma}$ is the Higgs expectation value
in unimodular gauge. As we have seen from our solutions in Fig.~3b, if no
radiation is generated, the quantity $h_{\gamma}$ oscillates at fixed
amplitude for cycle upon cycle, and so the integral (\ref{propertime})
diverges as $t$ is extended back into the past. If radiation \textit{is}
generated with each new cycle, the amplitude of $h_{\gamma}$ increases as we
follow the universe back into the past, as indicated in Fig.~4, and the
argument becomes even stronger. There are two possibilities: either
$h_{\gamma}$ remains finite, or it diverges at some finite value of conformal
time, $\tau_{*}$. From the action (\ref{cosmoAction}) one sees that it will do
so like $h_{\gamma}\propto1/(\tau-\tau_{*})$. In this case, the action for a
massive particle in Eq.~(\ref{propertime}) will diverge at $\tau_{*}$. Thus,
with this definition of geodesic completeness, we conclude that all cyclic
Higgs scenarios are geodesically complete to the past, whereas all
inflationary scenarios are not.

In sum, we have shown that the Higgs, if it is metastable, has profound
implications for cosmology. For big bang inflationary cosmology, metastability
is problematic: it is unexpected, requires highly improbable initial
conditions, and predicts a dire future in which the metastable phase ends and
the universe collapses in a big crunch. By contrast, metastability dovetails
with the cyclic picture for which decay of the current vacuum is a fundamental
prediction, required to end the current cycle and begin the next. Remarkably,
our Weyl-invariant formulation of the standard model has made it possible to
construct an action that incorporates all known microphysics and, at the same
time, has classical solutions that completely describe cyclic evolution, from
bounce to expansion to contraction to bounce again, with each cycle
reproducing similar physics that is like what we observe. We have further
demonstrated how the Higgs naturally makes the cyclic scenario geodesically complete.

We thank A. Ijjas for comments on the manuscript. Research at Perimeter
Institute is supported by the Government of Canada through Industry Canada and
by the Province of Ontario through the Ministry of Research and Innovation.
This research was partially supported by the U.S. Department of Energy under
grant number DE-FG03-84ER40168 (IB) and under grant number DE-FG02-91ER40671 (PJS).

\end{document}